\documentstyle[preprint,aps]{revtex}
\begin{document}
\draft
\date{\today}     
\title{The Energy of a Plasma in the Classical Limit}
\author{Merav Opher\footnote{email: merav@orion.iagusp.usp.br}, 
Reuven Opher\footnote{email: opher@orion.iagusp.usp.br}}
\address{Instituto Astron\^omico e Geof\' \i sico - IAG/USP, Av. Miguel 
St\'efano, 4200 \\ 
CEP 04301-904 S\~ao Paulo, S.P., Brazil}

\maketitle

\begin{abstract}
When $\lambda_{T} \ll d_{T}$, where $\lambda_{T}$ is the de Broglie 
wavelength and $d_{T}$, the distance of closest approach of thermal 
electrons, a classical analysis of the energy of a plasma can be made. 
In all the classical analysis made until now, 
it was assumed that the frequency of the fluctuations $\omega \ll T$ 
($k_{B}=\hbar=1$). Using the 
{\it fluctuation-dissipation theorem}, we evaluate 
the energy of a plasma, allowing the frequency of the fluctuations to be 
arbitrary. We find that the energy density is appreciably larger than 
previously thought for many interesting plasmas, such as the plasma 
of the Universe before the recombination era.

\end{abstract}

\section{Introduction}

There have been many classical calculations of the energy of a plasma 
\cite{ich,kra,corr}. They are based on perturbation theory of an ideal 
gas, in terms of the plasma parameter $g$ (which usually is a small 
value). The treatment, to the first order in $g$, is the Debye-H\"{u}ckel 
theory. However, in the calculations that have been made it is assumed that 
$\omega \ll T$ ($k_{B}=\hbar=1$). This is a very strong assumption. 
For example, in our previous analysis \cite{oph1,oph2}, 
we showed that only by not assuming $\omega \ll T$, 
is the blackbody spectrum obtained.

We evaluate the energy of a plasma, studying the electromagnetic fluctuations 
in a plasma without assuming that $\omega \ll T$.
A plasma in thermal equilibrium sustains fluctuations of the magnetic and 
electric fields. The electromagnetic fluctuations are described by the 
fluctuation-dissipation theorem \cite{sit}.

The evaluation of the electromagnetic fluctuations in a plasma has been made in 
numerous studies \cite{daw}. Recently, Cable and Tajima \cite{ct1} (see 
also \cite{ct2}) studied the magnetic field fluctuations in a cold plasma 
description with a constant collision frequency as well as for a warm, 
gaseous plasma, described by kinetic theory. 

Using a model that extends the work of Cable and Tajima \cite{ct1}, we study 
an electron-proton plasma of temperature 
$10^{4}-10^{5}~K$ with densities $10^{13}-10^{19}~cm^{-3}$. 
The condition for a classical analysis is that $\lambda_{T} < d_{T}$, 
where $\lambda_{T}$ is the de Broglie wavelength for a thermal electron 
and $d_{T}=e^{2}/T$, the distance of closest approach. This condition 
is satisfied for $T < 3\times 10^{5}~K$ and for the plasmas studied.

In section \ref{sec:flu} we recall the expressions for the electromagnetic 
fluctuations in a plasma, and in section \ref{sec:ene}, the 
electromagnetic energy is computed. Finally, we discuss our results 
in section \ref{sec:co}.
 
\section{Electromagnetic Fluctuations}
\label{sec:flu}

The spectra of the electromagnetic fluctuations in an isotropic plasma are 
given by \cite{sit},  
\begin{equation}
\frac{{\langle E^{2} \rangle}_{{\bf k}{\omega}}}{8\pi}= 
\frac{1}{e^{{\omega}/T}-1}\frac{Im~{\varepsilon}_{L}}
{{\mid {\varepsilon}_{L} \mid}^2}~~+~~
2\frac{1}{e^{{\omega}/T}-1}
\frac{Im~{\varepsilon}_{T}}
{{\mid {\varepsilon}_{T}-{\left( \frac{k c}{\omega}\right) }^{2} \mid}^2}~,
\label{ele}
\end{equation}
\begin{equation}
\frac{{\langle B^{2} \rangle}_{{\bf k}{\omega}}}{8\pi}=2
\frac{1}{e^{{\omega}/T}-1}
{\left( \frac{k c}{\omega} \right) }^{2}
\frac{Im~{\varepsilon}_{T}}
{{\mid {\varepsilon}_{T}-{\left( \frac{k c}{\omega}\right) }^{2} \mid}^2}
\label{magn}
\end{equation}
($\hbar=k_{B}=1$), where ${\varepsilon}_{L}$ and ${\varepsilon}_{T}$ are the 
longitudinal and transverse dielectric permittivities of the plasma.
The first and second terms of Eq. (\ref{ele}) are the 
longitudinal and transverse electric field fluctuations, respectively. 

By using the fluctuation-dissipation theorem, we can estimate the 
energy in the electromagnetic fluctuations for all frequencies and wave 
numbers. The calculation includes not only the energy of the fluctuations in 
the well defined modes of the plasma, such as 
plasmons in the longitudinal component and photons in the transverse 
component, but also the energy in fluctuations that do not propagate.

For the description of the plasma, we use the model described in 
detail in Opher and Opher \cite{oph1,oph2}. The description includes  
thermal and collisional effects. It uses the
equation of Vlasov in first order, with the BGK (Bhatnagar-Gross-Krook) 
collision term that is a model equation 
of the Boltzmann collision term \cite{cle}.  
We used the BGK collision term as a rough guide for the inclusion of 
collisions in a plasma.  
 	
From this description, the dielectric permittivities for an isotropic plasma 
are easily obtained:  
\begin{equation}
\varepsilon_{L}(\omega,{\bf k}) = 1 + \sum_{\alpha} 
\frac{{\omega_{p\alpha}}^{2}}{k^{2}{v_{\alpha}}^{2}}
\frac{1+\frac{(\omega+i\eta)}{\sqrt{2}kv_{\alpha}}Z \left ( 
\frac{\omega+i\eta_{\alpha}}{\sqrt{2}kv_{\alpha}} \right )}
{1+\frac{i\eta}{\sqrt{2}kv_{\alpha}}Z \left (
\frac{\omega+i\eta_{\alpha}}{\sqrt{2}kv_{\alpha}} \right )}~,
\label{el}
\end{equation}
\begin{equation}
\varepsilon_{T}(\omega,{\bf k}) = 1 + \sum_{\alpha} 
\frac{{\omega_{p\alpha}}^{2}}{\omega^{2}}
\left ( \frac{\omega}{\sqrt{2}kv_{\alpha}} \right ) Z \left ( 
\frac{\omega+i\eta_{\alpha}}{\sqrt{2}kv_{\alpha}} \right )~,
\label{et}
\end{equation}
where $\alpha$ is the label for the species of particles, $v_{\alpha}$ 
the thermal velocity for the species and $Z(z)$, the Fried and Conte 
function.

\section{Electromagnetic Energy}
\label{sec:ene}

In order to estimate the electromagnetic energy, we use the dielectric 
permittivities, given by Eqs. (\ref{el}) and 
(\ref{et}), and calculate the magnetic and the electric field 
spectra from Eqs. (\ref{ele}) and (\ref{magn}). 
Integrating the spectra in wave number and frequency 
(and dividing by ${(2\pi)}^{3}$), we obtain the energy densities of
the magnetic field $\rho_{B}$ and of the transverse and longitudinal 
electric fields $\rho_{E_{T}}$ and $\rho_{L}$. 

Usually, when estimating the energy stored in 
the electromagnetic fluctuations from Eqs. (\ref{ele}) and (\ref{magn}), 
it is assumed that $\omega \ll T$ ($k_{B}=\hbar=1$). 
With this assumption, the Kramers-Kronig relations can then be used, 
and a simple expression for the energy is obtained \cite{ich,kra}. 
However, the assumption that $\omega \ll T$ is very restrictive. 
For example, a large part of the fluctuations
which create the blackbody electromagnetic spectrum has
$\omega > T$ \cite{oph1,oph2}. It is therefore necessary to perform the 
integration of the spectra over frequency and wave number without using 
this assumption.

Our model uses kinetic theory with a collision term that describes the 
binary collisions in the plasma. A cutoff has to be taken since, for very 
small distances, the energy of the Coulomb interaction excedes the kinetic 
energy. This occurs for distances $r_{min} \sim e^{2}/T$, which defines 
our maximum wave number, $k_{max}$.

A large $k_{max}$ is needed in order to reproduce the blackbody
spectrum. In this study, we used a $k_{max}$ equal to the inverse of the 
distance of closest approach, which we previously found is able to do this 
\cite{oph1,oph2}. Any smaller $k_{max}$ was unable to reproduce the entire 
blackbody spectrum.  

In the usual classical calculations, the correction to the energy due to correlations 
between the particles, is made through the {\it correlation energy}. 
To the first order in the plasma parameter $g$, the correlation energy 
depends on the two-particle correlation function $S(k)$,
\begin{equation}
E_{C}=\frac{n}{4\pi^{2}}\int
dk k^{2} \phi_{k}S(k)-\frac{n}{4\pi^{2}}\int 
dk k^{2} \phi_{k}~,
\end{equation}
where the second term is the energy of the particles due to their 
own fields. $S(k)$ can be estimated by the fluctuation-dissipation theorem 
or by the BBGKY hierarchy equations \cite{corr}. 
Generally, it is assumed that $\omega \ll T$ (so the Kramers-Kronig relation 
can be used) and $S(k)$ is 
obtained as
\begin{equation}
S(k)= \frac{k^{2}}{k^{2}+k_{D}^{2}}~,
\end{equation}
where $k_{D}$ is the inverse of the Debye length.

With this, the energy density of a plasma to first order in $g$ is given as
\begin{equation}
U=\frac{3}{2}nT \left( 1-\left( \frac{g}{12\pi} \right) \right)~,
\label{corr}
\end{equation}
where $n$ is the number density of the particles. Thus, 
the correlation energy, to the first order in $g$ is
\begin{equation}
E_{c}=-\frac{3}{2}nT\left ( \frac{g}{12\pi} \right ) ~.
\end{equation}

We define the energy of a plasma as 
\begin{equation}
U=\frac{3}{2}nT (1+\Delta)~.
\end{equation}
With this definition, $\Delta=\Delta_{0}=-g/12\pi$, for 
the previous classical analysis (Eq. (7)), where the subscript ``0'' means 
that the assumption $\omega \ll T$ has been used.

Higher order calculations of the correlation energy have been made, 
for example by O'Neil and Rostoker \cite{or}. However, 
in all treatments, the assumption $\omega \ll T$ has been made. 
As we commented above, the assumption $\omega \ll T$ is very 
strong. A large part of the fluctuations has $\omega > T$. 

To obtain the interaction energy, we need to subtract 
the energy of the particles due to their own fields, the second term of 
Eq. (5),  from the longitudinal energy density, $\rho_{L}$. We thus have 
$\rho_{int}=\rho_{L}-\frac{n}{4\pi^{2}}\int dk k^{2} \phi_{k}$. 
Using Eq. (9), the interaction energy can be written as 
$\rho_{int} \equiv \frac{3}{2}(nT) \Delta$, where 
$\rho_{int}$ is the equivalent of the
correlation energy.  In fact, using the approximation $\omega \ll T$, 
$\rho_{int}$ is equal to the second term of Eq. (\ref{corr}).

In order to compare $\rho_{int}$ with $E_{c}$, 
we define the parameter,
\begin{equation}
F\equiv \frac{\mid \Delta \mid - \mid \Delta_{0} \mid}{\mid \Delta_{0} \mid}~.
\end{equation}

We previously found \cite{oph2} that the transverse energy (summing the 
transverse electric and magnetic field energies, $\rho_{E_{T}}$ and $\rho_{B}$) has 
an additional energy, compared to the blackbody energy density in vacuum. The 
additional transverse energy is
\begin{equation}
\Delta \rho_{\gamma} = \rho_{B}+\rho_{E_{T}}-\rho_{\gamma}~,
\label{delta}
\end{equation}
where $\rho_{\gamma}$ is the photon energy density, estimated 
as the blackbody energy density in vacuum. 

Adding the interaction energy $\rho_{int}$ to $\Delta \rho_{\gamma}$, we 
obtain the total change in the energy density due to the transverse
and longitudinal components,
\begin{equation}
\rho_{new} = \Delta \rho_{\gamma} + \rho_{int}~.
\label{new}
\end{equation}

We calculate $\rho_{new}$ and $\rho_{int}$ for an electron-proton 
plasma at $T=10^{5}~K$, $T=10^{4}~K$ and $T=10^{5}~K$ for densities 
ranging from $10^{3}-10^{19}~cm^{-3}$. The densities were 
chosen so as to assure that the plasma parameter, 
$g=1/n\lambda_{D}^{3} < 1$, in order that kinetic theory is valid. 
For these plasmas, the de Broglie wavelength is less than the distance of 
closest approach of thermal electrons, which justifies 
our classical treatment. 
 
In Figure 1, we plot $\Delta$ as a function of the density 
$10^{3}~cm^{-3} \leq n \leq 10^{19}~cm^{-3}$ for the temperatures 
$T=10^{3}~K$, $10^{4}~K$, and $10^{5}~K$. We extended each plot 
until the density for which $g=0.3$ was reached. For each of the temperatures, 
the value of $g$ increases with the density. In the case of $T=10^{5}~K$, 
for example, for $n=10^{3}~cm^{-3}$, $g=9.62\times 10^{-9}$ and for
$n=10^{9}~cm^{-3}$, $g=3.04\times 10^{-6}$. When $g=0.3$, 
$n=10^{19}~cm^{-3}$. In the case of $T=10^{3}~K$, for
$n=10^{3}~cm^{-3}$, $g=3.04\times 10^{-6}$ and for $n=10^{10}~cm^{-3}$, 
$g=9.62\times 10^{-3}$. When $g=0.3$, $n=10^{13}~cm^{-3}$. 

We found a very good fit for the results of Figure 1, using a Fermi-Dirac 
functional form for the density dependence of $\Delta$, 
$\Delta(T)=A1/(exp[(x/A2)-A3]+1)$, with $x=log(n)$ and 
$A1=a_{10}+a_{11}T+a_{12}T^{2}$; $A2=a_{20}+a_{21}T+a_{22}T^{2}$ and 
$A3=a_{30}+a_{31}T+a_{32}T^{2}$. From Figure 1, we obtain 
$A1=0.3522-0.1698(T/10^{5})+0.1145{(T/10^{5}) }^{2}$, 
$A2=0.8255+0.4797(T/10^{5})-0.4532{(T/10^{5}) }^{2}$ and 
$A3=17.650+33.027(T/10^{5})-26.201{(T/10^{5}) }^{2}$.
The curves (filled, dashed and dotted) are evaluated from the analytic 
expression; the filled squares are the calculated values of $\Delta$ from Eqs. (1)-(4). 
The fit can be seen to be excellent. 
In Figure 2, we plot $F=(\Delta-\Delta_{0})/\Delta_{0}$ as a function of 
the density, for the temperatures $T=10^{3}~K$, $10^{4}~K$ and $10^{5}~K$, 
which shows how $\Delta$ differs from the usual correction $\Delta_{0}$.

The values of $\Delta$ that we obtained are positive and larger in 
absolute value than $\Delta_{0}$, whereas $\Delta_{0}$ is negative. 
This indicates that the energy in the 
fluctuations dominates the interaction 
energy of the particles.  We observe that $F$ can reach values of a 
thousand or greater.

The additional transverse energy $\Delta \rho_{\gamma}$ is completely 
negligible for these temperatures and densities. For example, 
for $T=10^{5}~K$ and $n=10^{19}~cm^{-3}$, $\Delta \rho \cong 
10^{-3} \rho_{\gamma}$. For this temperature and density, 
$\rho_{par}=273\rho_{\gamma}$ and $\rho_{new}$ is 
completely dominated by $\rho_{int}=\Delta \rho_{par}$.
For example, for $T=10^{5}~K$ and $n=10^{17}~cm^{-3}$, 
$\Delta \rho_{\gamma} \cong 10^{-7} \rho_{\gamma}$.

As a check, we calculated $\rho_{int}$, integrating in frequency 
only up to $\omega=\omega_{p}$, 
the plasma frequency $(\ll T)$, and integrating in wavenumber up to 
$k \leq k_{D}$. As expected, we then found that $\Delta$ is equal to 
$\Delta_{0}$, the value obtained in previous analysis.

\section{Conclusions and Discussion}
\label{sec:co}

We calculated $\rho_{new}$ and $\rho_{int}$ for an electron-proton 
plasma as a function of density for $T=10^{3}-10^{5}~K$. For many interesting 
plasmas, we found that $\Delta \gg \Delta_{0}$. 
We used the BGK collison term as a rough guide to the inclusion 
of collisions. The BGK is a model collision term for 
the Boltzmann collision term. 
Collisions, however, change the results very little. For example, 
for $T=10^{5}~K$ and $n=10^{10}~cm^{-3}$, the difference in $\Delta $, 
with or without collisions, is less than $10^{-6}$. 
Since there is no significant difference between the energy 
density, with or without the collision term, the use of a more
acurate collision term than the BGK collision term is not necessary.

Appreciably different values than the usual ones are obtained, 
for the interaction energy of a plasma, by not assuming $\omega \ll T$. 
To the first order in $g$, we found that the energy of an ideal gas 
needs to be corrected by a positive value, approximately $0.3 \rho_{par}=
0.3 ({3}/{2})nT$. This results in very different values from the usual 
ones $\sim 10^{-3}-10^{-4}~ (3/2) nT$. 

We obtained a general expression for the correction $\Delta$ as a 
function of density and temperature: 
$\Delta(T)=A1/(exp[(x/A2)-A3]+1)$ with $x=log(n)$,  
$A1=0.3522-0.1698(T/10^{5})+0.1145{(T/10^{5}) }^{2}$, 
$A2=0.8255+0.4797(T/10^{5})-0.4532{(T/10^{5}) }^{2}$ and 
$A3=17.650+33.027(T/10^{5})-26.201{(T/10^{5}) }^{2}$.

The total correction to the energy is completely 
dominated by the interaction energy. For these temperatures and 
densities, the transverse additional energy is negligible. 
 
Our results may be applied to the plasma before the recombination era, 
when the plasma had a temperature 
$T > 10^{3}~K$ and a density $n >10^{3}~cm^{-3}$. Since the expansion rate of the Universe 
(the Hubble parameter) is proportional to the square root 
of the plasma energy density, our results indicate that the Universe before the 
recombination era was expanding appreciably faster than previously thought. 

The purpose of this work was to demonstrate that $\omega \ll T$ is an 
extremely strong assumption. By not making this assumption, there is a large 
change in the energy of the plasma.

The authors would like to thank the anonymous referees for helpful comments. 
M.O. would like to thank the Brazilian agency FAPESP
for support (no. 97/13427-8) and R.O. the Brazilian agency CNPq for 
partial support. Both authors would like to thank the Brazilian project 
Pronex/FINEP (no. 41.96.0908.00) for support.

\begin{figure}
\caption{The correction $\Delta$ as a function 
of density and temperature. The filled curve is for $T=10^{5}~K$, 
the dashed curve for $T=10^{4}~K$ and the dotted curve for $T=10^{3}$. 
The curves are evaluated from the analytic expression and the filled 
squares are the calculated values from Eqs. (1)-(4).} 
\label{fig1}
\end{figure}
%%%%%%%%%%%%%%%%%%%%%%
%%%%%%%%%%%%%%%%%%%%%%
\begin{figure}
\caption{The deviation of the correction, $\Delta$ from 
the usual one, $\Delta_{0}$: $F=\mid \Delta \mid -
\mid \Delta_{0} \mid /\mid \Delta_{0} \mid.$ 
The filled curve is for $T=10^{5}~K$, 
the dashed curve for $T=10^{4}~K$ and the dotted curve for $T=10^{3}$.
}
\label{fig2}
\end{figure}
%%%%%%%%%%%%%%%%%%%%%%

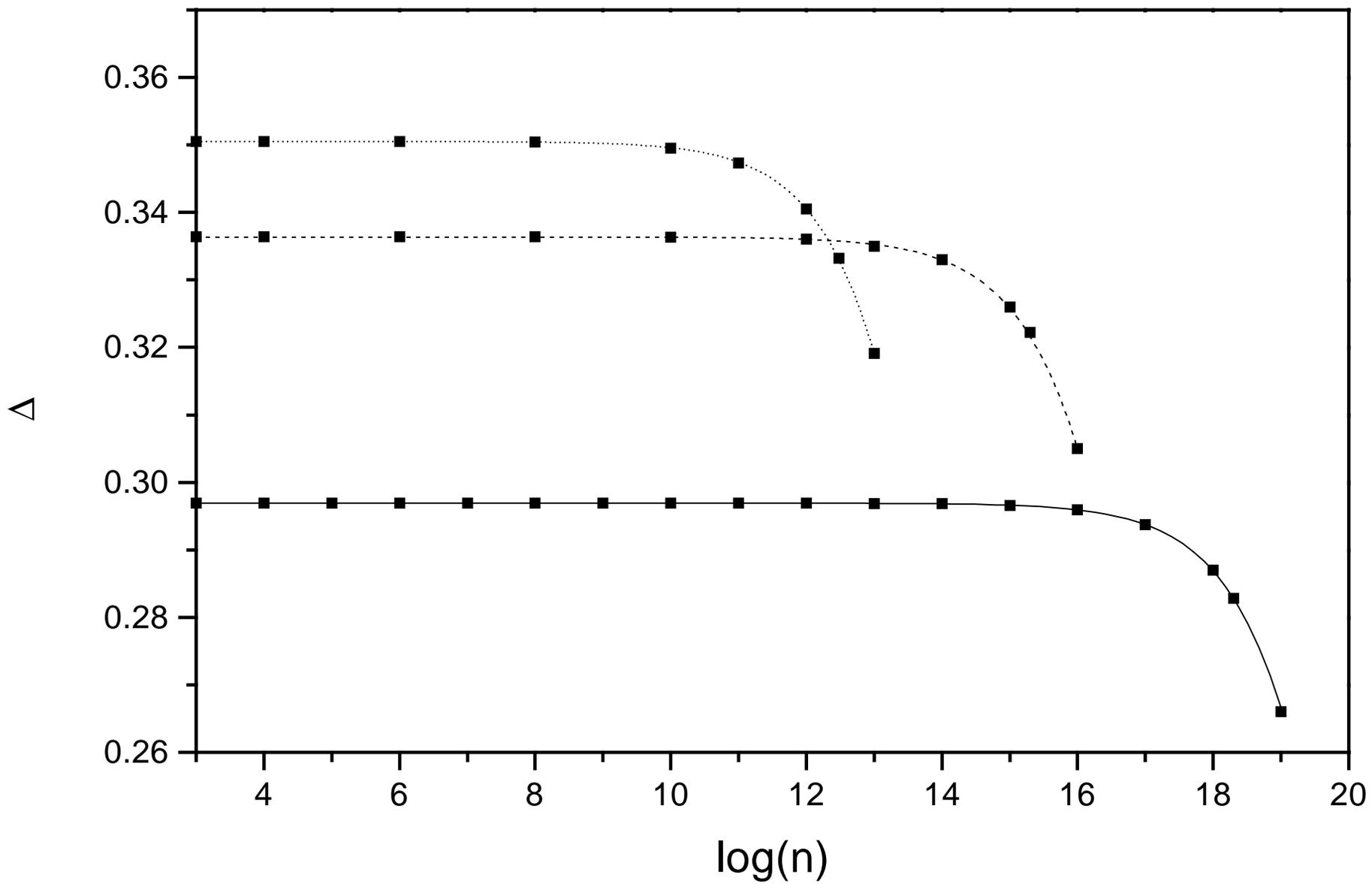
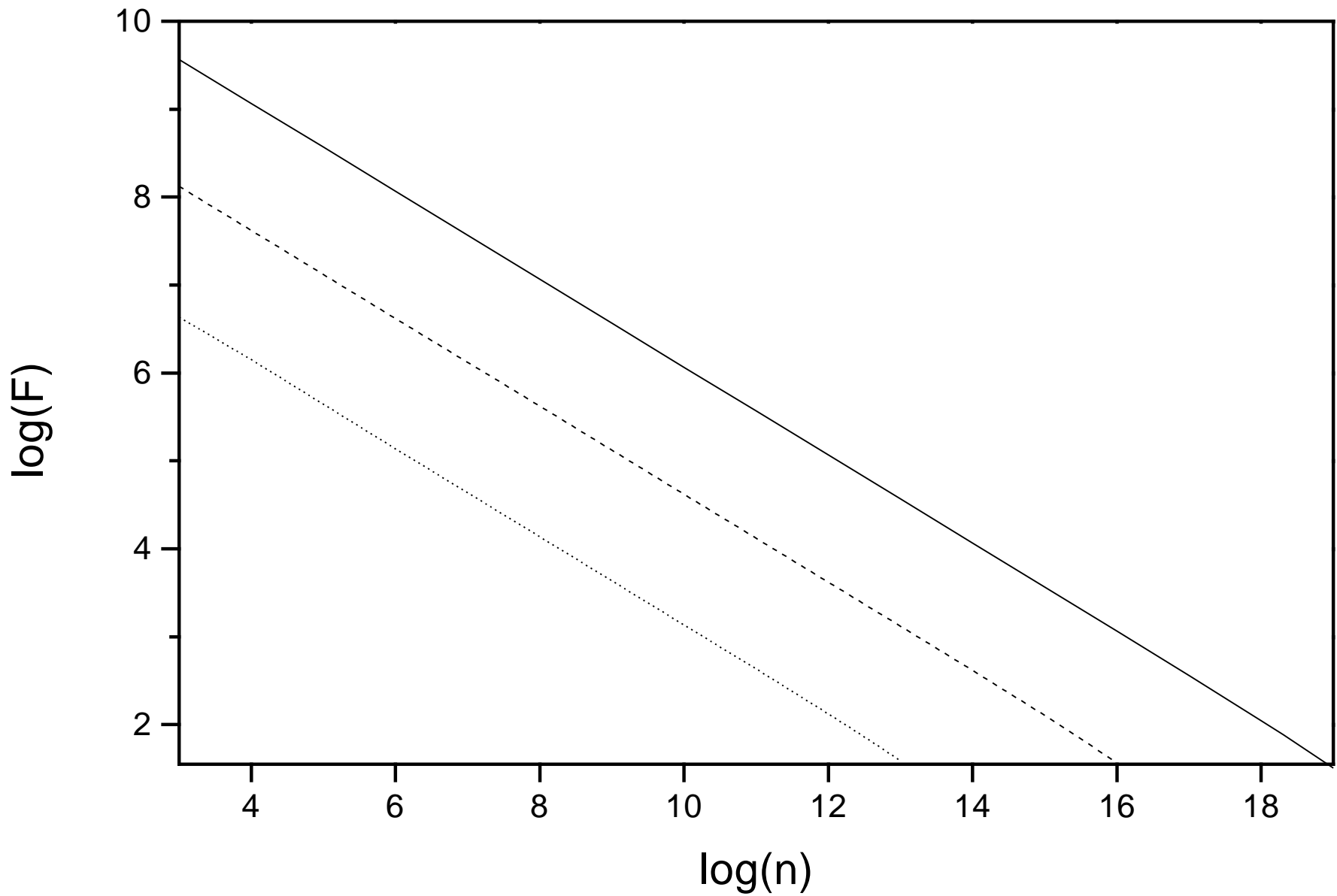
\end{document}